\documentclass[12pt,times,twocolumn]{article}

\usepackage{sbcm}

\usepackage{graphicx,url}

 \usepackage[utf8]{inputenc}
 \usepackage{fancyvrb}
 \usepackage{fancyhdr}
 \fancypagestyle{firststyle}
 {
	 \fancyhf[C]{16th Brazilian Symposium on Computer Music, SBCM 2017}
	       \fancyfoot{}
	       }

\title{Vivace: a collaborative live coding language and platform}

\author{Vilson Vieira\inst{1}\and Guilherme Lunhani\and\\ Geraldo Magela de Castro
  Rocha Junior\and Caleb Mascarenhas Luporini\and\\ Daniel Penalva\and Ricardo
  Fabbri\inst{2}\and Renato Fabbri\inst{3}}
\address{Cod.ai \\
	SC, Brazil
         \nextinstitute
         Polytechnic Institute -- IPRJ/UERJ \\
	 Rua Bonfim, 25 - Vila Amélia --
	 CEP 28625-570, Nova Friburgo, RJ, Brazil
         \nextinstitute
	 Visualization, Imaging and Computer Graphics lab --
         VICG/ICMC/USP \\
	 Av. Trabalhador São-Carlense, 400 - Centro --
	 CEP 13566-590, São Carlos, SP, Brazil
         \email{vilson@void.cc, rfabbri@gmail.com, renato.fabbri@gmail.com}
}
\begin{document}

\maketitle

\begin{abstract}
Live coding is a performance and creative technique based on improvised and interactive coding.
Many recent endeavors have focused in live coding both because of aesthetics
and as a way to alleviate performance drawbacks when the musical instrument is a computer.
This paper describes the principles and the design of Vivace, a live
coding language and environment built with Web technologies to be
executed on web browsers.
The approach is compelling by 1) allowing many performers to code simultaneously;
2) the synthesis of audio and video;
3) a very simple syntax;
4) being a multiplatform software.
We also strive to contextualize Vivace by means
of historical and usage summaries including a live coding sub-genre. 
\end{abstract}
\thispagestyle{firststyle}
\section{Introduction and narrative}
Live coding is an artistic performance and creative technique
based upon writing software code in a live and improvised manner~\cite{nilson2007live}.
It can be used to generate e.g. sound, images, video, lights and poetry
although it is prevalent in computer music~\cite{eff}.
Most often, the code is continually changed and projected in a large surface as a way to
make it visible to the audience~\cite{collins2011live}.
The usage of live coding in non-performative contexts is also reported,
such as in sound design and art installations~\cite{eff}.
In this paper, we describe Vivace, a live coding language and platform.
It emerged from pragmatic and aesthetic needs, as described in the following sections.
The software is cross-platform because it is based on web technologies,
such as HTML5 Video and Web Audio API, and is oriented towards video and music rendering.
The language is very simple, allowing for a primary goal of
Vivace to be achieved: the simultaneous writing of the code by many performers
and the audience.

In summary, this paper describes how
a number of artistic presentations motivated the creation of the language and platform,
describes the Vivace language and platform and how this endeavor
lead to the emergence of a live coding sub-genre
called ``freak coding'' (e.g. by its manifesto~\cite{freak} and related artists).

\subsection{A historical outline} 
In November of 2011, a live coding trio called
\textit{FooBarBaz}~\cite{foobarbaz} unleashed its first presentation
for a wide audience. Its performers used two instances of
ChucK~\cite{wang2003chuck} and a dedicated mixing instance composed by Puredata and an analog
mixer~\footnote{Pictures of the presentation available at
\url{http://www.flickr.com/photos/festivalcontato/6436260557}}.
  Live coding had been gaining world wide popularity~\cite{nilson2007live,
  collins2003live, brown2007a, collins2011live} which motivated the creation of a dedicated
  congress, the International Conference on Live Coding (ICLC)~\cite{iclc},
  already on its third edition, and a network of
  public events and movements like Algorave~\cite{algorave}. Live Coding has been adopted by
  performers also in Brazil, like Guilherme Lunhani, Antonio Goulart, André
  Damião Bandeira and Magno Caliman. However, when considering massive
  audience, live coding practice remains quite untouched in Brazil.
To the best of our knowledge, the presentation
was the first live coding performance in our country with such a wide audience -- almost 5,000
attendees were in the gathering where code was used on-the-fly to
create the music they were listening.
At the same time, two live coding desktop work-spaces were projected on large screens to the
public, following the principles of the \textsc{toplap} manifesto~\cite{ward2004live}.

During the performance, the trio used ChucK in an unconventional
way. Instead of writing loops and conditionals, one of the live coders
manipulated parameters of audio files by editing lists of numerical
values together with mnemonic operations like retrograde and
transposition. The other live coder focused on more fluid lines with
large sounds with evolving characteristics; this contributed for coherent
musical arcs. Audio mixing with Puredata was carried out by
the third performer literally using handwaving gestures tracked by a
camera and custom color detection algorithms designed by us. Live coders
used code templates for quick insertion in the text editors (Vi
and Emacs).
Other visual resources the performers focused on:
Unix ``cowsay'' generated phrases and animated bouncing balls
-- stimulating the audience to imitate Rapid Eyes
Movements (REM) -- on both terminals, i.e. on both screens projected to the audience.
The performance was reported as interesting by technicians,
artists and the general audience.
Nonetheless, it was altogether complex, not to say messy.

Based on the aforementioned elements, and the need for greater simplicity
and interactivity with the public, the Vivace was designed as a new live coding language and platform
~\footnote{Live demonstrations of Vivace are on-line at
    \url{http://void.cc/freakcoding} and~\url{http://void.cc/cranio},
    ready to be used by everyone using
  Google Chrome, Mozilla Firefox or Apple Safari.}~\cite{vivace}. To avoid software
configuration, and to make it easy to share the session and the
system, the Web was chosen as the running
environment for Vivace. On every new session performed using Vivace,
new principles were added into the language and, at the same time,
into our artistic approach.

\subsection{Additional motivation \& inspiration: arrange the room, the code is dirty}

\begin{figure*}[htpb]
  \begin{center}
    \includegraphics[scale=.35]{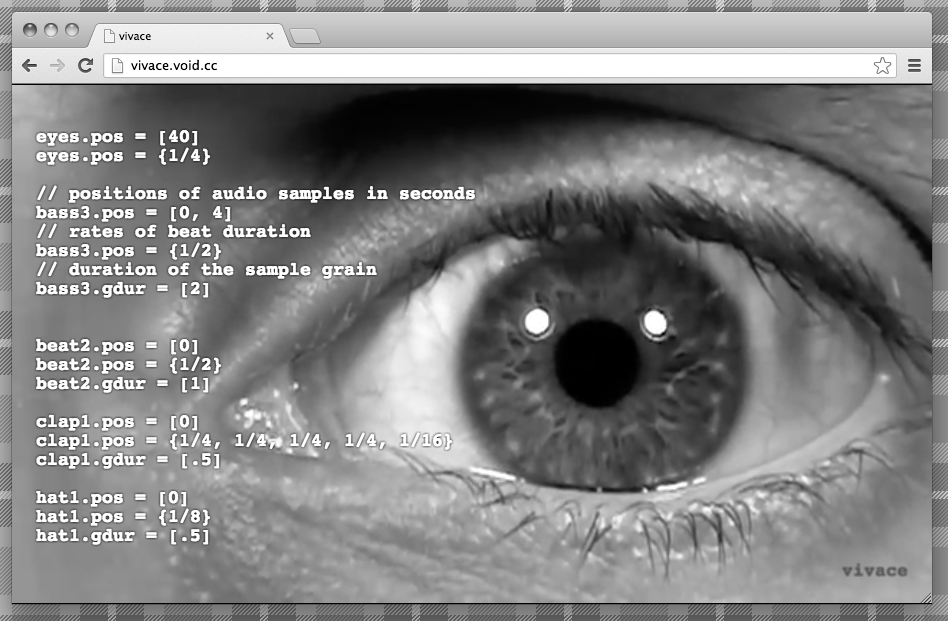}
    \caption{The Vivace platform basic interface: video playback is in the background while the code is in the foreground and controls both video and musical audio. The interface is accessed as a usual HTML page in a web browser.}
    \label{fig:vivace}
  \end{center}
\end{figure*}

Vivace is inspired by various live coding languages. The syntax of
Vivace, as shown in Figure~\ref{fig:vivace}, borrows elements from ixi
lang~\cite{magnusson2011ixi} such as the use of sequences to control
audio parameters in real time. ABT~\cite{fabbri} and
FIGGUS~\cite{fabbri2} were tightly relevant to the development of
Vivace as well and we are planning to rewrite some of their components
-- originally in Python -- inside Vivace.
ChucK was an influence from the beginning, and Vivace resembles the simplified
interfaces that we constructed in ChucK which were often minimized into lists in few lines.
Fluxus~\cite{fluxus} was also inspirational for the Vivace
environment where the code is shown on top of the video frames.

After the creation of the Audio Data API~\cite{audiodata} and the most
recent Web Audio API~\cite{webaudio} -- easing real time audio
processing inside Web browsers -- a collection of audio Web
applications started to emerge. The same holds for live coding
languages and systems.  Thus, in addition to the \textit{desktop
  languages} above, Vivace was also inspired by recent Web
live coding languages and is part of this \textit{family} of Web
applications together with Gibber~\cite{gibber},
livecoder~\cite{livecoder}, livecoding.io~\cite{livecodingio}, 
livecodelab~\cite{livecodelab} and Wavepot~\cite{wavepot} to mention just a few.

A remarkable difference between Vivace and other languages and
environments in the same family is the element of
collaborativity. Vivace was built to enable writing code by many
hands at the same time, as with the now popular ``e-pads'' or
collaborative real-time text editors~\footnote{Etherpad on:
  \url{http://etherpad.org}.}, a feature which is naturally
implementable on the Web. Another difference is the unconcern to be a
Turing-complete language. This made the design of Vivace more flexible
and closer to musical thinking as opposed to a computing process (a
characteristic perceived in ixi lang as well). Vivace is designed to
associate the precision of code and the flexibility of artistic
expression while maintaining simplicity.

\section{The language (specification) we all speak}

Vivace, as a language~\footnote{The complete specification can be
  found at
  \url{https://github.com/automata/vivace/wiki/Language-spec} together with the
  grammar (\url{https://github.com/automata/vivace/blob/master/vivace.jison})
  and lexical rules
  (\url{https://github.com/automata/vivace/blob/master/vivace.jisonlex}) specified in Bison and Flex dialects,
  respectively.}, is a
collaborative live coding language with use of extremely simple
syntax, mnemonic operations, easy audio mixing, template editing and
audio parameters automation. The use of shared code, sounds and images
leads to a more complex scenario, thus increasing the possibility of
inconsistency of compiled code as well as artistic
results.

Vivace is not an imperative language. Instead of routines and
procedures to control audio attributes, it uses definitions related
with musical scores and the \emph{track paradigm} common on music
production software~\cite{collins2011live}. It is natural to musicians (and, as we
experienced during performances, also to non-musicians) to understand
a sequence of notes, or audio parameters, repeating over and over
again, than for-loops and if-chains. In this way, Vivace is a
declarative, domain specific language, based on the following
principles:

\begin{itemize}
	\item Names are literals like $foo$, $bar$, $baz$ and are defined as
	  the user wants.
	\item Music is constituted by voices (instruments).
	\item Voices have name, timbre and parameters changing along time.
	\item The language should be simple. One only defines some properties with a set of
	  values (i.e. arrays, dictionaries) making it possible to generate
	  sequences.
	\item Mnemonic musical operations (reverse, inverse, transpose) on
	  properties by use of syntax sugar: few chars, powerful changes.
	\item Timbre are signals made by chains of audio generators and
	  filters or video files as described below.
	\item Parameters are musical notes, amplitudes, oscillator
	  frequencies, delay time and so on.
	\item Parameters change their values at specific times and for certain
  durations.
\end{itemize}

Here is a ``Hello, World!'' Vivace code~\footnote{YOUTUBE\_URL
stands for any youtube video url such as \url{http://www.youtube.com/watch?v=XXX}}:

\begin{Verbatim}[fontfamily=courier, xleftmargin=\parindent,fontsize=\scriptsize]
# foo is a simple audio sample,
# oscillator or video file
foo.src = youtube('YOUTUBE_URL')
# defining the video positions
# (in seconds) to be played
foo.pos = [10, 20, 35]
# the durations,
# as time ratios of a pulse,
# to be played at each position:
foo.cdur = [1/2, 1/4, 1/8, 1/16, 1]
\end{Verbatim}

A voice is defined as \textit{foo} and its parameters are specified
using the \textit{dot} operator. Every parameter changes over time as
the values written in numerical sequences, surrounded by brackets. A
special sequence exists to every parameter. This is essentially all of
the Vivace syntax.

There are extra semantics to operate on the sequences. Every sequence
accepts operators: mnemonic commands used to reverse, transpose and
even replace elements of the sequence based on list
comprehensions. Those operations are common in music composition~\cite{collins2011live} and
having them as mnemonics makes typing fast and handy for live
coding. The next listing presents the standard operators:

\begin{Verbatim}[fontfamily=courier, xleftmargin=\parindent,fontsize=\scriptsize]
# one can use operators
foo.pos = [1, 2, 3] reverse
# result is [3, 2, 1]
foo.pos = [1, 2, 3] inverse 
# result is [1, 0, -1]
foo.pos = [1, 2, 3] transpose +2
# result is [3, 4, 5]

# list comprehension
foo.pos = [i+1 for i in [1, 2, 3]] 
# result is [2, 3, 4]

# or combine both
foo.pos=[i+1 for i in [1, 2]] reverse 
# result is [3, 2] as expected
\end{Verbatim}

Vivace is written in JavaScript to take advantage of Web technologies.
To parse Vivace, Jison~\cite{jison} comes handy, a JavaScript
library that clones Flex and Bison functionality as lexer and
parser. This flexibility to parse and execute new languages as
JavaScript inside every browser opens a remarkable opportunity to
experiment with new syntax and semantics for live coding. To make the
Vivace editor collaborative we used ShareJS~\cite{sharejs} which makes
Web applications content live concurrent. ShareJS uses a multiple clients and
one server network architecture. By running a common server for both
Vivace Web application and files, it is possible
to share Vivace code with any client accessing a common
URL. Furthermore, considering the tradition of UI design and
development on the Web thanks to HTML and CSS, one can experiment
those new languages with fast prototyped UI -- a requisite already
addressed by live coding languages~\cite{mclean2010visualisation,
  magnusson2011algorithms} like Texture~\cite{mclean2011texture},
Al-Jazari and Betablocker. Along these advantages, it is important to
note: every live coding language built on the Web runs everywhere a
browser is installed. No firewall chain to bypass for OSC, no software
installation and configuration, no dependencies, people just need to
type an URL.

\subsection{Vivace audio and video engine}
Before the Web Audio API, the only way to create sound in web pages
was using plug-ins. Recently, the Web Audio API enabled real time
audio processing on Web browsers~\footnote{At the time this paper was
  written, more than 74\% of current available Web browsers support Web Audio
  API~\cite{caniuse}, including the most populars Google Chrome, Mozilla Firefox and Apple
  Safari.}. Every functionality is implemented as native code (in C++
and Assembly when appropriate) to guarantee maximum performance. The
API is based on a convenient and familiar paradigm: audio unit
graphs. Web Audio specifies a collection of nodes (\emph{AudioNode}
objects) and routines to connect and disconnect them. While
manipulating those nodes we can create a large number of audio
applications: synthesizers, filters, analyzers, mixers and even real
time audio engines for live coding. This motivated basic explorations
of multichannel expansion, filtering and audio effects, controlling an
integrated Web audio system.

\begin{figure}[htpb]
    \begin{center}
      \includegraphics[scale=.5]{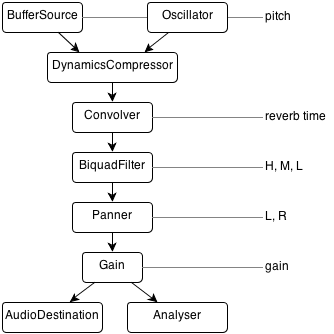}
      \scriptsize\caption{\label{fig:chain} Standard audio processing units are directly related to Web Audio API objects for
        each voice within Vivace.}\par
    \end{center}
  \end{figure}

\begin{figure}[htpb]
    \begin{center}
      \includegraphics[scale=.2]{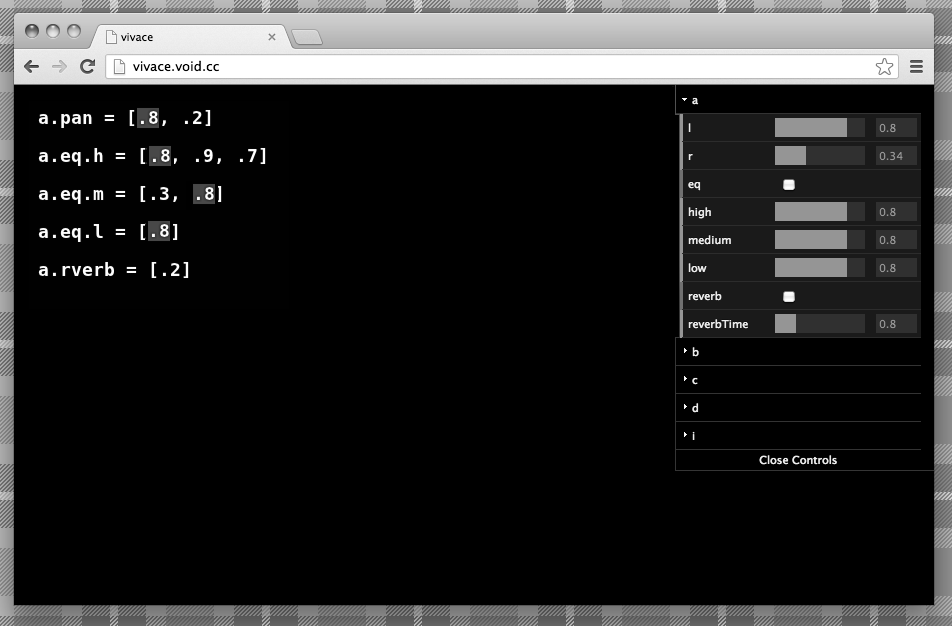}
      \caption{Every audio unit parameter can be manipulated by code or
        using the UI in the Vivace platform.}
      \label{fig:ui}
    \end{center}
\end{figure}

Every voice in Vivace is represented as a default audio chain such as
the one shown in Figure~\ref{fig:chain}. All audio unit parameters
within this chain (e.g. pitch, reverb time, high, medium and low
channel levels, panner values and gain) can be manipulated editing the
code or by sliders on a GUI (Figure ~\ref{fig:ui}).

This kind of
interface is more familiar to musicians, resembling a real mixer, and
enables an adequate treatment of voice timbre and spatialization of
the sound sources by means of parameters like level of stereophonic
channels L and R, quality, central frequency and gain of a 3-band
equalization filter, and reverb time control.

Vivace supports every audio unit implemented by Web Audio API. It is
possible to load audio files or synthesize in real time using
wave-table oscillators. The default audio chain of each voice can be
modified at any time while it is running. It is interesting to note
the presence of an ``Analyzer'' inside the default chain. It uses FFT
(natively implemented) to expose energies and frequencies, enabling
the use of those values to animate videos and render graphical forms
inside Vivace.

Along with audio, Vivace supports video files. It is possible to
upload files or use YouTube URLs. Videos are treated the same way as
buffer sources or oscillators, i.e.\ as voices, and can be manipulated
in real time, making Vivace a live cinema or a VJing tool.

\section{Into the wild: the rise of freak coding makes it collaborative}
\begin{figure}[htpb]
  \begin{center}
    \includegraphics[scale=.2]{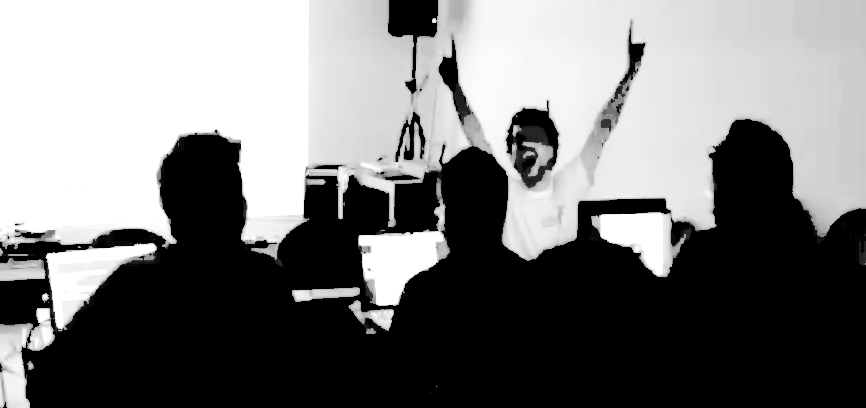}
    \caption{Freak coder: a live coder who uses
      popular and ``freak'' media to boost the attention of the audience to the shared
      code.}
    \label{fig:freakcoder}
  \end{center}
\end{figure}

Vivace as a tool enables interaction while everyone can use their
own creativity. The interaction is not mediated by a common score, but
by a mutual desire to create a composition in real time. In this
context the \emph{freak coder} was born (Figure~\ref{fig:freakcoder}):
someone that adds his individuality with others, aiming to transform
the computer into an instrument of artistic fruition, without
restricting to himself the control of the machine but inviting
everyone to join him in the activity. A freak coder decides what
he is going to do and amplifies his own comprehension of the computer
capacity as an instrument. By using simple rules, Vivace enables the
emergence of the performance and makes it a kind of a collective game,
where the rules, being visible to everyone through the code, eases audience and
specialists alike to join in.

Live coding becomes a natural path to the type of use and
technological development in which freak coders are involved, in
confluence with the understanding that technology should never be
treated as a dogma or kept in secret. Live coding is seen as a
behavioral de-alienation of a digital artist.
The triad
performer, code and audience characterizes the performance as live
coding.  This comprehension was possible after a presentation by
labMacambira.sourceforge.net at the $9^{th}$ edition of AVAV
(\textit{\'{A}udioVisual Ao Vivo} or Live Audiovisual), an event where
artists who are experimenting with audio and video in real time come
together to show their works. In this presentation, the authors Caleb
Luporini and Gera Rocha started without Renato Fabbri and Vilson
Vieira, as they were on their way to the presentation, traveling from
another city. Upon arrival, Fabbri and Vilson turned their laptops on and
started taking part on the performance in such a way that no
embarrassment or rupture was brought into the event.  It is important
to state that no previous rehearsal had taken place between
Mr. Luporini and Mr. Rocha.\footnote{Videos were selected beforehand
  by Mr. Luporini alone without knowledge of the other performers.} In
the 30 minutes-long performance, the audience started to take guidance
from messages given on the performance large screen and actually edited
Vivace code that was being played together with the starting four
performers.

Another artifact noted on the presentation was the emergence, in a
formal environment~\footnote{The four performers were in a light-less
  room, three of them facing the big screen and the other one facing
  the public.}, of a collective euphoria fertilized by a human-machine
interaction. It is the performer's posture that takes a spectator to
an experience of a non-spectator and to take part on a highly
technological activity as something playful and possible to be
assimilated. During the entire presentation, all
labMacambira.sourceforge.net members were cheerful and established a
relation of lightness and brotherhood with the audience.  Spectators
were being constantly invited by the posture of
labMacambira.sourceforge.net members to interact with what was being
proposed.  This interplay between the four elements therein present --
performers, computer, Vivace and audience -- created an
environment of collaboration and liberty as generators of playfulness
and technical knowledge unheard of, at least in Brazilian live coding,
to our knowledge. This is the ``facilitator'' that emerged and received the name
\emph{freak coder}.

\begin{figure}[htpb]
  \begin{center}
    \includegraphics[scale=.2]{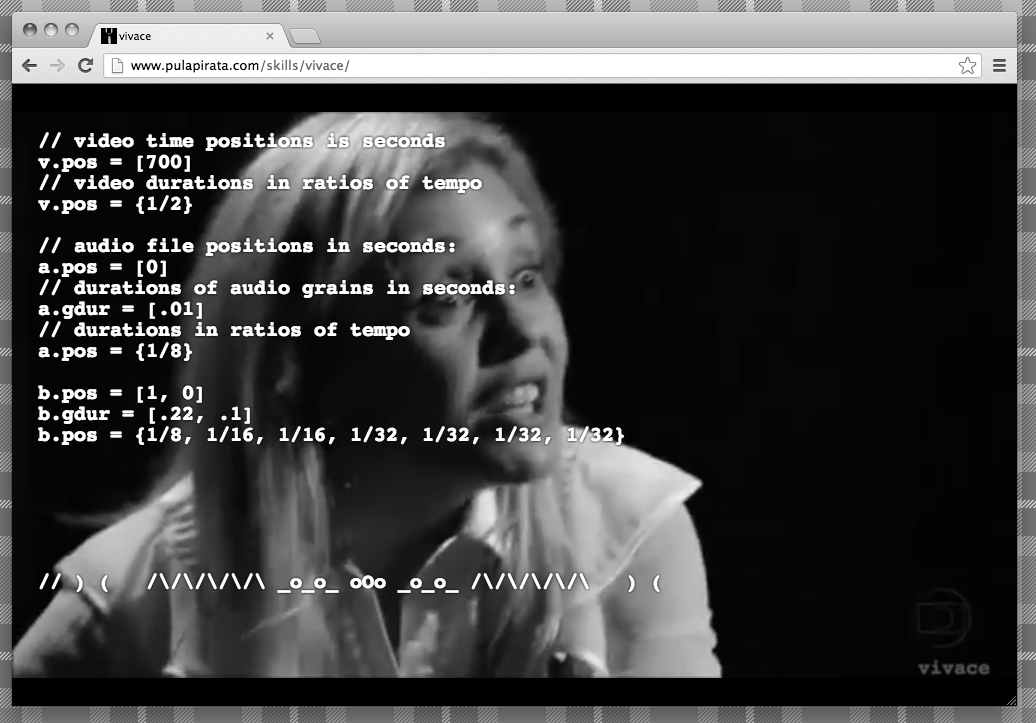}
    \caption{Videos sampled from popular Brazilian novels and B-movies
      were used as material to attract the attention of the audience to the code
	  in the ``freak-coding'' aesthetics and live coding sub-genre.}
    \label{fig:novela}
  \end{center}
\end{figure}

To attract the attention of the wider audience, we as freak coders
used popular media as material. The code was displayed in front of
video scenes sampled from popular Brazilian novels (as in
Figure~\ref{fig:novela}) and B-movies, which resulted in a ``freak'' style,
with images of monsters and funny dialogues between novel actors. In
other performances for heterogeneous attendees the effect was the same
as the first presentation where we used these kinds of pop-art: the
people was fascinated by the adherence between the code and the media
they see every day on their TV sets. Since then, the use of popular and
``freak'' media has become a signature of ``freak coding''.

All labMacambira.sourceforge.net members take part in the Brazilian
free software movement. In a way, freak coding origins should be
looked for inside this movement. It is inherent to the free software
movement the continued transmission of what is known. The same happens
on the demystification of technology and the festive and gregarious
behavior. At the performer-computer relation is where this behavior
becomes concrete. More than the materials used in the live coding
sessions, the performer's stance in relation to the computer -- as
already expressed in the described presentation -- is what really
subverts not just the highly technical computer use but the
relationship between the human and the machine. Namely a kind of a
``rock and roll'' stance. The freak coder breaks, by his own nature,
the stigma of the computer as the source of a serious and professional
posture. In the same way, breaks with the posture of the scholar
performer, stern and closed in herself. The freak coding is ``rock and
roll''. The freak coder becomes Jerry Lee of technology making
``techno-pyrophagy''. He codes and cheers at same time. The freak
coder seduces through the computer screen and by the way he codes.

\section{Conclusions and future work}
Vivace was motivated by actual performances that took place in the
recently emerging Brazilian live coding scene. The development of the
language was guided by this direct contact of performers and the wider
public. The language was designed and implemented after the
identification of common patterns already used on
presentations and the need for simplicity and interactivity.
Following open source practice, Vivace is
developed by many hands from computer scientists, musicians, activists
and social scientists. At present, the language is certainly not perfect nor
all-encompassing, but it does strike a useful balance between
flexibility and rigor, making it an interesting language for artistic
expression on collaborative sessions.

It is important to note the advantage of using the Web as the platform
for experimentation on live coding and other computer music
approaches. Recent APIs like Web Audio together with the rapid
prototyping of multi-platform UI and language parsers creates a
prolific scenario. Henceforth the most interesting characteristic is
the collaboration proportioned by the Web. Using collaborative editors
we can expose an entire music program to be edited by anyone,
anywhere.

Vivace, although a ``freak coding'' language, is constrained in its
music expression. Having a domain specific language as Vivace is
interesting to express some musical ideas where it is hard or
even impossible elsewhere. In this context we assume Vivace as
one of the many tools and a contribution to create other languages and
collaborative systems emerging from live coding practices. In this
way, we can tell that the described performances and even Vivace are
motivating the creation of other live coding
tools. Carnaval~\footnote{Carnaval is being conceived as free software
  and a collaborative art piece since its beginning. The first
  sketches are on-line at \url{http://automata.cc/carnaval}} is one of
these new realizations, it can be seen as a ``personal TV
channel''. Each channel is related to a Vivace instance, making it
possible for anyone to remix media and create their own
composition. It is a social network of live coded remixes. Vivace,
instead of an isolated piece of software is then used as a module,
a part of Carnaval.

In our experiences as performers and developers of live coding
languages we can assert this style of music realization as
inspirational and flexible. Nevertheless, we continue to search for
improvements on Vivace -- and others derived tools -- to increase an
already consolidated objective of live coding as a musical practice:
make computer music performance more human, more interactive with the
wider audience~\cite{collins2011live}.

Future improvements are planned on Vivace: the possibility to
explicitly define large musical arcs as nested sequences related to
audio units, the use of 3D graphics APIs to render forms -- and relate them to running audio parameters -- 
and text messages to the audience, and improved UI to make the code editing more
flexible and reactive~\cite{brett}. Along with the language and
system itself, this paper is a live initiative. Freak coding as an
artistic style (a sub-genre of live coding) will be explored more deeply on future studies --
regarding its aesthetics -- and in already planned performances.

We want to end by underlining the importance of social aspects
regarding live coding. The authors
were not working physically close to each other since the beginning,
i.e. since the creation of
Vivace and the rise of freak coding. The performances emerged from
the gathering of the artists and programmers,
which gave rise to new tools and
aesthetics.
On the other hand,
given the continuous audiovisual feedback the audience and performers
have from the code,
we noticed that the potential of live coding for
computer programming demystification and introduction
is often emphasized by enthusiasts and the literature~\cite{eff}.

\bibliographystyle{unsrt}
\bibliography{./cmjbib}

\end{document}